\documentstyle[graphicx,aps,multicol]{revtex}
\draft

\begin{document}

\title{Radiative strength functions in $^{93-98}$Mo}
\author{M.~Guttormsen,$^{1}$\footnote{Electronic address: magne.guttormsen@fys.uio.no} R.~Chankova,$^{1}$ U.~Agvaanluvsan,$^{2,3}$ E.~Algin,$^{2,3,4,5}$ L.A.~Bernstein,$^{2}$ F.~Ingebretsen,$^{1}$ T.~L{\"o}nnroth,$^{6}$ S.~Messelt,$^{1}$ G.E.~Mitchell,$^{3,4}$ J.~Rekstad,$^{1}$ A.~Schiller,$^{2}$ S.~Siem,$^{1}$ A.C.~Sunde,$^{1}$ A.~Voinov,$^{7}$ and S.~{\O}deg{\aa}rd$^{1}$}
\address{$^1$ Department of Physics, University of Oslo, N-0316 Oslo, Norway}
\address{$^2$ Lawrence Livermore National Laboratory, L-414, 7000 East Avenue, Livermore, CA 94551, USA}
\address{$^3$ North Carolina State University, Raleigh, NC 27695, USA}
\address{$^4$ Triangle Universities Nuclear Laboratory, Durham, NC 27708, USA}
\address{$^5$ Department of Physics, Osmangazi University, Meselik, Eskisehir, 26480 Turkey}
\address{$^6$ Department of Physics, {\AA}bo Akademi, FIN-20500 Turku, Finland}
\address{$^7$ Department of Physics and Astronomy, Ohio University, Athens, Ohio 45701, USA}
\maketitle

\begin{abstract}
Radiative strength functions (RSFs) in $^{93-98}$Mo have been extracted using the ($^3$He,$\alpha \gamma$) and ($^3$He,$^3$He$^{\prime} \gamma$) reactions. The RSFs are U-shaped as function of $\gamma$ energy with a minimum at around $E_{\gamma}=3$ MeV. The minimum values increase with neutron number due to the increase in the low-energy tail of the giant electric dipole resonance with nuclear deformation. The unexpected strong increase in strength below $E_{\gamma}=3$ MeV, here called soft pole, is established for all $^{93-98}$Mo isotopes. The soft pole is present at all initial excitation energies in the $5-8$ MeV region.
\end{abstract}


\begin{multicols}{2}

\section{Introduction}

The $\gamma$-decay of nuclei at high excitation energy tends to follow certain statistical rules. The dominating $\gamma$-transition driving factors are the number of accessible final states and the $\gamma$-ray transmission coefficient. The largest uncertainties are connected to the latter factor.  In the description of this factor Blatt and Weisskopf \cite{blatt} included an $E_{\gamma}^{2L+1}$ dependency, where $L$ is the angular momentum transfer in the transition. In their definition of the radiative strength function (RSF), this simple energy dependence was divided out. With such a definition, the single-particle RSF (Weisskopf) estimates become independent of $\gamma$-ray energy. Various concepts of RSFs and $\gamma$ decay in the continuum are outlined in the reviews of Bartholomew {\sl et al.} \cite{bart1,bart2}.

It has been well known that the RSF is not at all constant, but shows an additional $E_{\gamma}^x$ dependency with $x=1-2$ for $\gamma$ energies in the $4-8$ MeV region. Axel \cite{axel} argued that this feature is due to the collective giant electric dipole resonance (GEDR), which represents the essential mechanism for the $\gamma$-decay. However, the situation is more complex. Further studies \cite{joly,gutt84,igash} reveal fine structures in the RSF, which are commonly called pygmy resonances. This name does not refer to specific structures: the E1 pygmy resonance in the $E_{\gamma}=5-7$ MeV region of gold to lead nuclei could be due to neutron skin oscillations \cite{enders00}, whereas bumps in the 3 MeV region of rare earth nuclei are now determined to be of M1 character \cite{schi04,krti}. The electromagnetic character and measured strength of the latter pygmy resonance is compatible with the scissors mode. Recently \cite{emel0,voin04}, the RSF picture of iron isotopes has been further modified by the observation of an anomalous increase in strength at $\gamma$ energies below 4 MeV. It is clear that the present situation urges for new experimental results.

The stable molybdenum isotopes are well suited as targets for the study of nuclear properties when going from spherical to deformed shapes. In this work we perform a systematic analysis of the RSFs of the six $^{93-98}$Mo isotopes. The RSFs depend on the dynamic properties of electric charges present within these systems ($Z=42$). Since the nuclear deformation varies from spherical shapes ($\beta \sim 0$) at $N= 51$ to deformed shapes ($\beta \sim 0.2$) at $N=56$, we expect to observe effects due to shape changes. Furthermore, these nuclei reveal weak GEDR tails at low $E_{\gamma}$ making them interesting objects in the search for other weak structures in the RSF.

The Oslo Cyclotron group has developed a sensitive tool to investigate RSFs for $E_{\gamma}$ below the neutron binding energy $S_n$. The method is based on the extraction of primary $\gamma$-ray spectra at various initial excitation energies $E_i$ measured in particle reactions with one and only one charged ejectile. From such a set of primary $\gamma$-spectra, nuclear level densities and RSFs can be extracted \cite{hend1,melb0,schi0}. The level density reveals essential nuclear structure information such as thermodynamic properties and pair correlations as function of temperature. These aspects of the molybdenum isotopes will be presented in a forthcoming work. Various applications of the Oslo method have been described in Refs.~\cite{schi1,gutt1,schi2,gutt3,voin1,siem1,dy03,femo}.

\section{Experimental method}

The particle-$\gamma$ coincidence experiments were carried out at the Oslo Cyclotron Laboratory for $^{93-98}$Mo using the CACTUS multi-detector array. The charged ejectiles were detected with eight particle telescopes placed at an angle of 45$^{\circ}$ relative to the beam direction. An array of 28 NaI $\gamma$-ray detectors with a total efficiency of $\sim$15\% surrounded the target and particle detectors. 

In the present work, results from eight different reactions on four different targets are discussed. Results from two of those reactions have been reported earlier. The beam energies for the different reactions are given in parentheses:\\
1) $^{94}$Mo($^3$He,$\alpha \gamma$)$^{93}$Mo (new, 30 MeV)\\
2) $^{94}$Mo($^3$He,$^3$He$^{\prime}\gamma$)$^{94}$Mo (new, 30 MeV)\\
3) $^{96}$Mo($^3$He,$\alpha \gamma$)$^{95}$Mo (new, 30 MeV)\\
4) $^{96}$Mo($^3$He,$^3$He$^{\prime}\gamma$)$^{96}$Mo (new, 30 MeV)\\ 
5) $^{97}$Mo($^3$He,$\alpha \gamma$)$^{96}$Mo (reported in \cite{emel0,femo}, 45 MeV)\\
6) $^{97}$Mo($^3$He,$^3$He$^{\prime}\gamma$)$^{97}$Mo (reported in \cite{emel0,femo}, 45 MeV)\\
7) $^{98}$Mo($^3$He,$\alpha \gamma$)$^{97}$Mo (new, 45 MeV)\\
8) $^{98}$Mo($^3$He,$^3$He$^{\prime}\gamma$)$^{98}$Mo (new, 45 MeV)\\

The targets were self-supporting metal foils enriched to $\sim 95$\% with thicknesses of $\sim 2$ mg/cm$^2$. The experiments were run with beam currents of $\sim 2$ nA for 1--2 weeks. The reaction spin windows are typically $I\sim 2-6 \hbar$.

The experimental extraction procedure and the assumptions made are described in Refs.~\cite{hend1,schi0} and references therein. For each initial excitation energy $E_i$, determined from the ejectile energy and reaction $Q$ value, $\gamma$-ray spectra are recorded. Then the spectra are unfolded using the known $\gamma$-ray response function of the CACTUS array \cite{unfold}. These unfolded spectra are the basis for making the first generation (or primary) $\gamma$-ray matrix \cite{gutt0}, which is factorized according to the Brink-Axel hypothesis \cite{axel,brink} as 
\begin{equation}
P(E_i,E_{\gamma}) \propto  \rho (E_i -E_{\gamma}) {\mathcal{T}}  (E_{\gamma}).
\label{eq:axel}
\end{equation}
Here, $\rho$ is the level density and ${\mathcal{T}}$ is the radiative transmission coefficient. 

The $\rho$ and ${\mathcal{T}}$ functions can be determined by an iterative procedure \cite{schi0} through the adjustment of each data point of these two functions until a global $\chi^2$ minimum of the fit to the experimental $P(E_i,E_{\gamma})$ matrix is reached. It has been shown \cite{schi0} that if one solution for the multiplicative functions $\rho$ and ${\mathcal{T}} $ is known, one may construct an infinite number of other functions, which give identical fits to the $P$ matrix by
\begin{eqnarray}
\tilde{\rho}(E_i-E_\gamma)&=&A\exp[\alpha(E_i-E_\gamma)]\,\rho(E_i-E_\gamma),
\label{eq:array1}\\
\tilde{{\mathcal{T}}}(E_\gamma)&=&B\exp(\alpha E_\gamma){\mathcal{T}} (E_\gamma).
\label{eq:array2}
\end{eqnarray}
Consequently, neither the slope ($\alpha$) nor the absolute values of the two functions ($A$ and $B$) can be obtained through the fitting procedure.

The parameters $A$ and $\alpha$ can be determined by normalizing the level density to the number of known discrete levels at low excitation energy~\cite{ENSDF} and to the level density estimated from neutron-resonance spacing data at the neutron binding energy $S_n$~\cite{IA98}. The procedure for extracting the total level density $\rho$ from the resonance energy spacing $D$ is described in Ref.~\cite{schi0}. Here, we will only discuss the determination of parameter $B$ of Eq.~(\ref{eq:array2}), which gives
the absolute normalization of ${\mathcal{T}}$. For this purpose we utilize experimental data on the average total radiative width of neutron resonances at $S_n$ $\langle\Gamma_{\gamma} \rangle$. 

We assume here that the $\gamma$-decay in the continuum is dominated by E1 and M1 transitions. For initial spin $I$ and parity $\pi$ at $S_n$, the width can be written in terms of the transmission coefficient by~\cite{kopecky} 
\begin{eqnarray}
\langle\Gamma_\gamma\rangle=\frac{1}{2\rho(S_n, I, \pi)} \sum_{I_f}&&\int_0^{S_n}{\mathrm{d}}E_{\gamma} B{\mathcal{T}}(E_{\gamma})
\nonumber\\
&&\rho(S_n-E_{\gamma}, I_f),
\label{eq:norm}
\end{eqnarray}
where the summation and integration run over all final levels with spin $I_f$, which are accessible by $\gamma$ radiation with energy $E_{\gamma}$ and multipolarity E1 or M1. 

\begin{figure}
\includegraphics[totalheight=17.5cm,angle=0,bb=0 0 350 730]{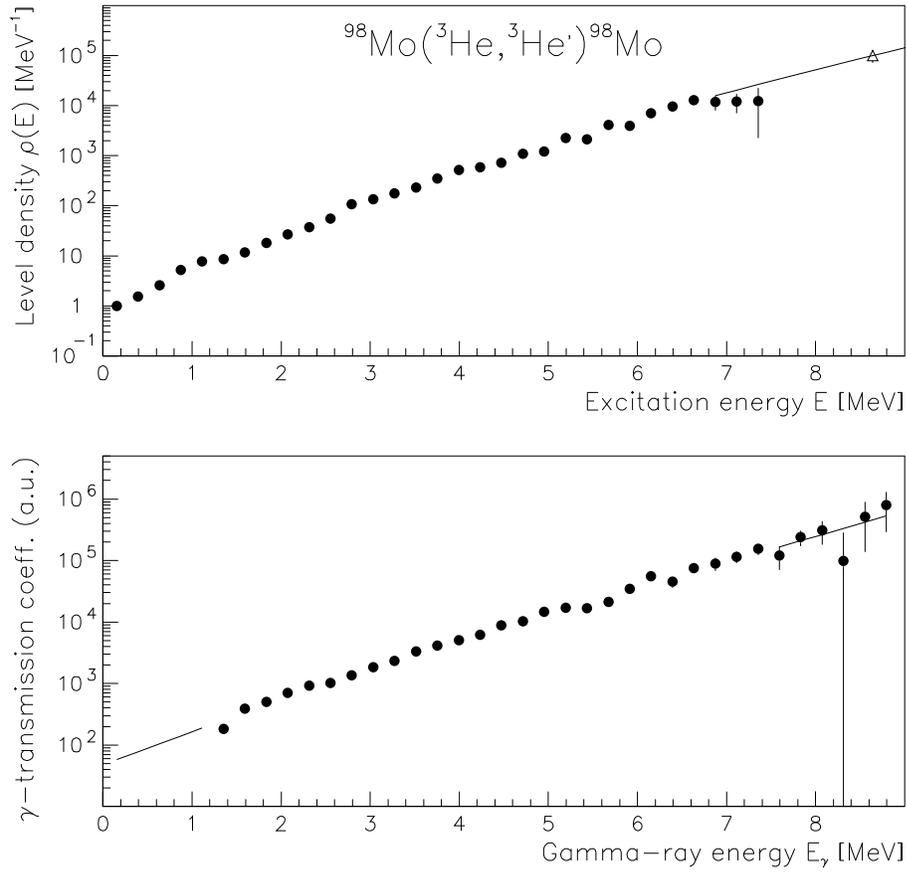}
\caption{Measured level density $\rho$ (upper panel) and radiative transmission coefficient ${\cal T}$ (lower panel) for $^{98}$Mo. The straight lines are extrapolations needed to calculate the normalization integral of Eq.~(\ref{eq:norm}). The triangle in the upper panel is based on resonance spacing data at $S_n$.}
\label{fig:rhorsf}
\end{figure}

A few considerations have to be made before $B$ can be determined. Methodical difficulties in the primary $\gamma$-ray extraction prevents determination of the functions ${\mathcal{T}}(E_{\gamma})$ in the interval $E_{\gamma}<1$~MeV and $\rho(E)$ in the interval $E > S_n-1$~MeV. In addition, ${\mathcal{T}}(E_{\gamma})$ at the highest $\gamma$-energies, above $E_{\gamma} \sim S_n-1$~MeV, suffers from poor statistics. For the extrapolation of $\rho$ we apply the back-shifted Fermi gas level density as demonstrated in Ref.~\cite{dy03}. For the extrapolations of ${\mathcal{T}}$ we use an exponential form. As a typical example, the extrapolations for $^{98}$Mo are shown in Fig.~\ref{fig:rhorsf}. The contribution of the extrapolations of $\rho$ and ${\cal T}$ to the calculated radiative width in Eq.~(\ref{eq:norm}) does not exceed $15$\% \cite{voin1}. The experimental widths $\langle\Gamma_\gamma\rangle$ in Eq.~(\ref{eq:norm}) are listed in Table~\ref{tab:tab1}. For $^{94}$Mo, this width is unknown and is estimated by an extrapolation based on the $^{96}$Mo and $^{98}$Mo values.  

The total radiative strength function for dipole radiation ($L=1$) can be calculated from the normalized transmission coefficient ${\mathcal{T}}$ by
\begin{equation}
f (E_{\gamma}) =\frac{1}{2\pi} \frac{ {\mathcal{T}}(E_{\gamma})}{ E_{\gamma}^3}.
\label{eq:fT}
\end{equation}
The RSFs extracted from the eight reactions are displayed in Fig.~\ref{fig:webrsf}. As expected, the RSFs seem not to show any clear odd-even mass differences. The results obtained for the ($^3$He,$\alpha$) and ($^3$He,$^3$He$^{\prime}$) reactions populating the same residual nucleus reveal very similar RSFs. Also for $^{96}$Mo two different beam energies have been applied, giving very similar RSFs. Thus, the observed energy and reaction independency gives further confidence in the Oslo method.

\section{Description of the radiative strength functions}

\begin{figure}
\includegraphics[totalheight=17.5cm,angle=0,bb=0 0 350 900]{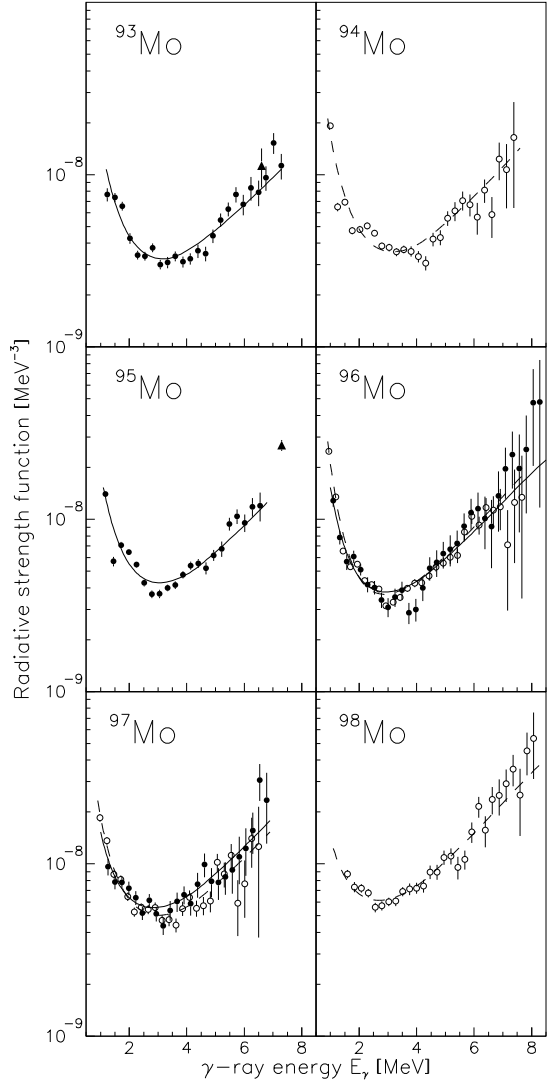}
\caption{Normalized RSFs for $^{93-98}$Mo. The filled and open circles represent data taken with the ($^3$He,$\alpha$) and ($^3$He,$^3$He$^{\prime}$) reactions, respectively. The filled triangles in $^{93,95}$Mo are estimates of $E1$ RSF of hard primary $\gamma$-rays \protect\cite{KU95} . The solid and dashed lines are fits to the RSF data from the two respective reactions (see text).}
\label{fig:webrsf}
\end{figure}

An inspection of the experimental RSFs of Fig.~\ref{fig:webrsf} reveals that the RSFs are increasing functions of $\gamma$ energy for $E_{\gamma} > 3$ MeV. This indicates that the RSFs are 
influenced by the tails of the giant resonances. As follows from 
previous work, the main contribution (about 80\%) is due to 
the electric dipole resonance (GEDR). The magnetic resonance 
(GMDR) and the isoscalar E2 resonance are also present in this region.


If the GEDR is described by a Lorentzian function, one will find that the strength function approaches zero in the limit $E_\gamma\rightarrow0$.
However, the $^{144}$Nd(n,$\gamma \alpha )$ reaction 
\cite{Popov} strongly suggests that 
$f_{{\mathrm{E}}1}$ has a finit value in this limit. Kadmenski{\u{\i}}, Markushev and Furman (KMF) 
have developed a model~\cite{kad} describing this feature 
for the electric dipole RSF:
 

\begin{equation} 
f_{\mathrm{E1}}(E_\gamma,T)=\frac{1}{3\pi^2\hbar^2c^2} \frac{0.7\sigma_{\mathrm{E1}}\Gamma_{\mathrm{E1}}^2(E_\gamma^2+4\pi^2T^2)} {E_{\mathrm{E1}}(E_\gamma^2-E_{\mathrm{E1}}^2)^2}.
\label{eq:E1}
\end{equation}
The temperature $T$ depends on the final state $f$ and for simplicity
we adapt the schematic form
\begin{equation}
T(E_f)=\sqrt{U_f/a},
\label{eq:tef}
\end{equation}
where the level density parameter is parametrized as $a=0.21 A^{0.87}$ MeV$^{-1}$. The intrinsic energy is estimated by $U_f=E_f-C_1-E_{\rm pair}$ with a back-shift parameter of $C_1=-6.6A^{-0.32}$ MeV ~\cite{egidy}. The pairing energy contribution $E_{\rm pair}$ is evaluated from the three point mass formula of Ref.~\cite{doba}. 


Although this model has been developed for spherical nuclei, it 
has been successfully applied to $^{56,57}$Fe and several rare 
earth nuclei \cite{voin04,voin1,siem1,dy03} assuming a 
constant temperature parameter $T$ in Eq.~(\ref{eq:E1}), i.e. one that is independent of
excitation energy. 
  In this work we assume that the temperature depends on excitation 
  energy according to Eq.~(\ref{eq:tef}), which gives an increase in the RSF at low $\gamma$ energy.

In order to compare the excitation energy dependent RSF with 
experiments, Eq.~(\ref{eq:E1}) should be folded with the 
final excitation energies of the specific experiment 
giving 
\begin{equation}
\langle f_{\mathrm{E1}}(E_\gamma) \rangle=\frac{1}{E_2-E_1}\int_{E_1-E_{\gamma}}^{E_2-E_{\gamma}}{\mathrm{d}}E_f f_{\mathrm{E1}}(E_f),
\label{eq:fE1ave}
\end{equation}
where the integration runs over all final excitation energies $E_f$ which are experimentally accessible by transition  with given energy $E_{\gamma}$. Our data set of primary $\gamma$-ray spectra is typically taken between $E_1 \sim 4$ MeV and $E_2 \sim S_n$ initial excitation energy.

The M1 GMDR contribution to the total RSF is described by a Lorentzian form 
through
\begin{equation}
f_{\mathrm{M1}}(E_\gamma)=\frac{1}{3\pi^2\hbar^2c^2} \frac{\sigma_{\mathrm{M1}}E_\gamma\Gamma_{\mathrm{M1}}^2} {(E_\gamma^2-E_{\mathrm{M1}}^2)^2+E_\gamma^2\Gamma_{\mathrm{M1}}^2}.
\label{eq:M1}
\end{equation}
This approach is in accordance with numerous experimental data 
obtained so far, and is recommended in Ref.\cite{IA98}.

Although of minor importance,  
the E2 radiation strength $f_{{\rm E}2}$ has also been included. Here, we use the same Lorentzian function as in Eq.~(\ref{eq:M1}), but with different resonance parameters and an additional factor $3/(5E_{\gamma}^2)$. The resonance parameters for the E1, M1 and E2 resonances are taken from the compilation of Refs.~\cite{IA98,atom} and are listed in Table~\ref{tab:tab1}.
 
The enhanced RSF at low $\gamma$ energies has at present no theoretical explanation. Recently, the same enhancement has been observed in the iron isotopes \cite{emel0,voin04}. We call this structure a soft pole in the RSF and choose a simple power law parametrization given by
\begin{equation}
f_{\rm soft pole}=\frac{1}{3\pi^2\hbar^2c^2}{\cal{A}}E_{\gamma}^{-b},
\end{equation}
where $\cal{A}$ and $b$ are fit parameters, and $E_{\gamma}$ is given in MeV.

Previously, a pygmy resonance around $E_{\gamma}\sim 3$ MeV has been reported in several rare earth nuclei~\cite{voin1,siem1,dy03}. The electromagnetic character of the corresponding RSF structure is now established to be of M1 type \cite{schi04,krti} and is interpreted as the scissors mode. Deformed nuclei can in principle possess this collective motion, and e.g.,~$^{98}$Mo with a deformation of $\beta \sim 0.18$, could eventually show some reminiscence of the scissors mode. Data on $^{94}$Mo \cite{frans03} and $^{96}$Mo \cite{frans04} show a summed $M1$ strength to mixed symmetry $1^+$ states around $\sim 3.2$ MeV in the order of $\sim 0.6 \mu_N^2$. This is about one order of magnitude lower than the $M1$ strength observed in well-deformed rare earth nuclei using the present method. This $M1$ strength is deemed too weak to cause a visible bump in our RSFs above 3 MeV.

We conclude that a reasonable composition of the total RSF is
\begin{equation}
f=\kappa (f_{{\mathrm{E}}1} + f_{{\mathrm{M}}1} +f_{\rm soft pole})+E_{\gamma}^2f_{\mathrm{E}2},
\end{equation}
where $\kappa$ is a normalization constant. Generally, its value deviates from unity for several reasons; the most important reasons are theoretical uncertainties in the KMF model and the evaluation of $B$ in Eq.~(\ref{eq:norm}). We use $\kappa$, ${\cal {A}}$ and $b$ as free parameters in the fitting procedure, and the results for the eight reactions are summarized in Table~\ref{tab:tab2}.

\begin{figure}
\includegraphics[totalheight=17.5cm,angle=0,bb=0 0 350 730]{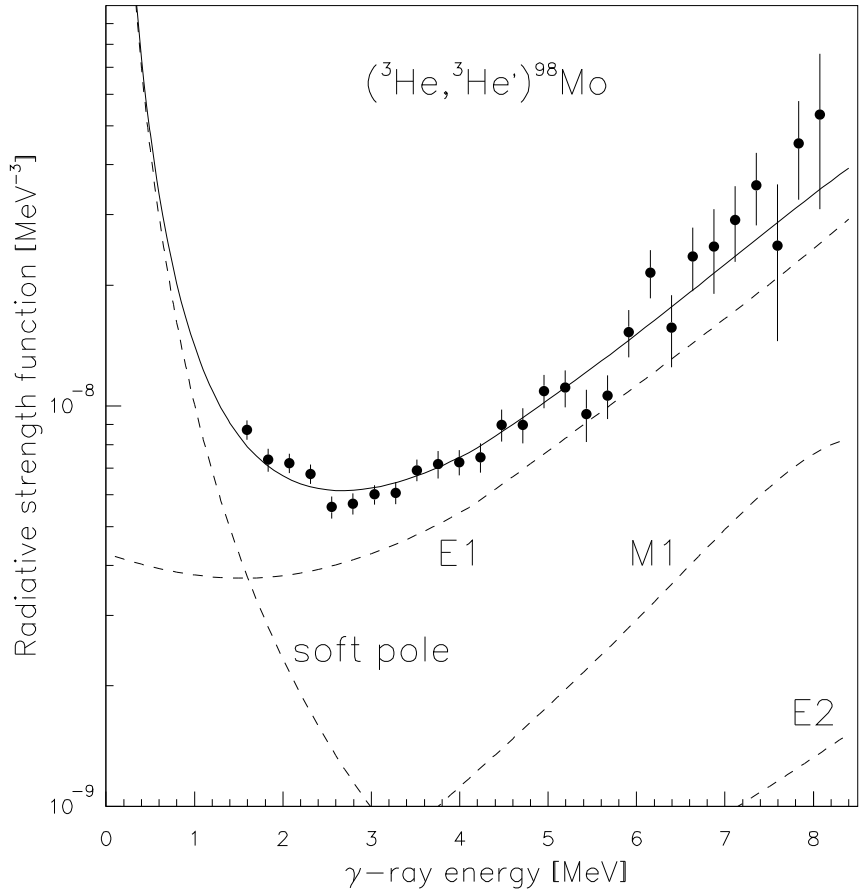}
\caption{Experimental radiative strength function of $^{98}$Mo compared to a model description including GEDR, GMDR and the isoscalar E2 resonance. The empirical soft pole component is used to describe the low energy part of the RSF.}
\label{fig:fitexp}
\end{figure}

In Fig.~\ref{fig:fitexp} the various contributions to the total RSF of $^{98}$Mo are shown. The main components are the GEDR resonance and the unknown low energy structure. We observe that the E1 component exhibits an increased yield for the lowest $\gamma$ energies due to the increase in temperature $T$. However, this effect is not strong enough to explain the low energy upbend. 

Figure~\ref{fig:webrsf} shows the fit functions for all reactions and gives qualitative good agreements with the experimental data. The fitting parameters $\kappa$, ${\cal{A}}$ and $b$ are all similar within the uncertainties. It should be noted that the soft pole parameters coincide with the description of the $^{57}$Fe nucleus~\cite{voin04} having ${\cal{A}}= 0.47 (7)$ mb/MeV, and $b=2.3(2)$. 

The almost a factor-of-two higher values of the RSFs for $E_{\gamma} > 3$ MeV when going from $N=51$ to $56$ can be understood from the corresponding evolution of nuclear deformation. Following the onset of deformation the GEDR will split into two parts, where $1/3$ of its strength is shifted down in energy and $2/3$ up. Photoneutron cross sections \cite{atom} do not show a splitting into two separate bumps, however, the observed increase in width $\Gamma_{{\rm E}1}$ as function of neutron number (see Table~\ref{tab:tab1}) supports the idea of a splitting, which is well known from other more deformed nuclei. Figure~\ref{fig:webrsf} demonstrates that the adopted widths describe very well the variation of the RSF strength as function of mass number.

In order to test that the prominent soft pole structure is present in the whole excitation energy region, we have performed the following test. Assuming that the level density from Eq.~(\ref{eq:axel}) is correct, we can estimate the shape of the strength functions starting at various initial excitation energies using

\begin{equation}
f(E_\gamma,E_i)=\frac{1}{2\pi} \frac{{\cal N}(E_i)P(E_i,E_{\gamma})}{\rho(E_i-E_{\gamma})E_{\gamma}^3}.
\label{eq:fei}
\end{equation}
Actually, $f(E_\gamma,E_f)$ would have been the proper expression to investigate, but due to technical reasons we choose $f(E_\gamma,E_i)$. One problem is that the normalization constant is only roughly known through the estimate

\begin{equation}
{\cal N}(E_i)=\frac{\int_0^{E_i} {\mathrm{d}}E_{\gamma }\rho(E_i-E_{\gamma})
{\cal T}(E_{\gamma})}{\int_0^{E_i}{\mathrm{d}}E_{\gamma}P(E_i,E_{\gamma})}
\label{eq:nei}
\end{equation}
with $E_i<S_n$. However, for the expression $f(E_\gamma,E_i)$ we are only interested in the shape of the RSFs, and an exact normalization is therefore not crucial. The evaluation assumes that eventual temperature dependent behavior of the RSF is small compared to the soft pole structure\footnote{Simulations using the KMF model indicate a maximum 20 \% effect from temperature dependence of the RSF.}.

\begin{figure}
\includegraphics[totalheight=21cm,angle=0,bb=0 0 350 900]{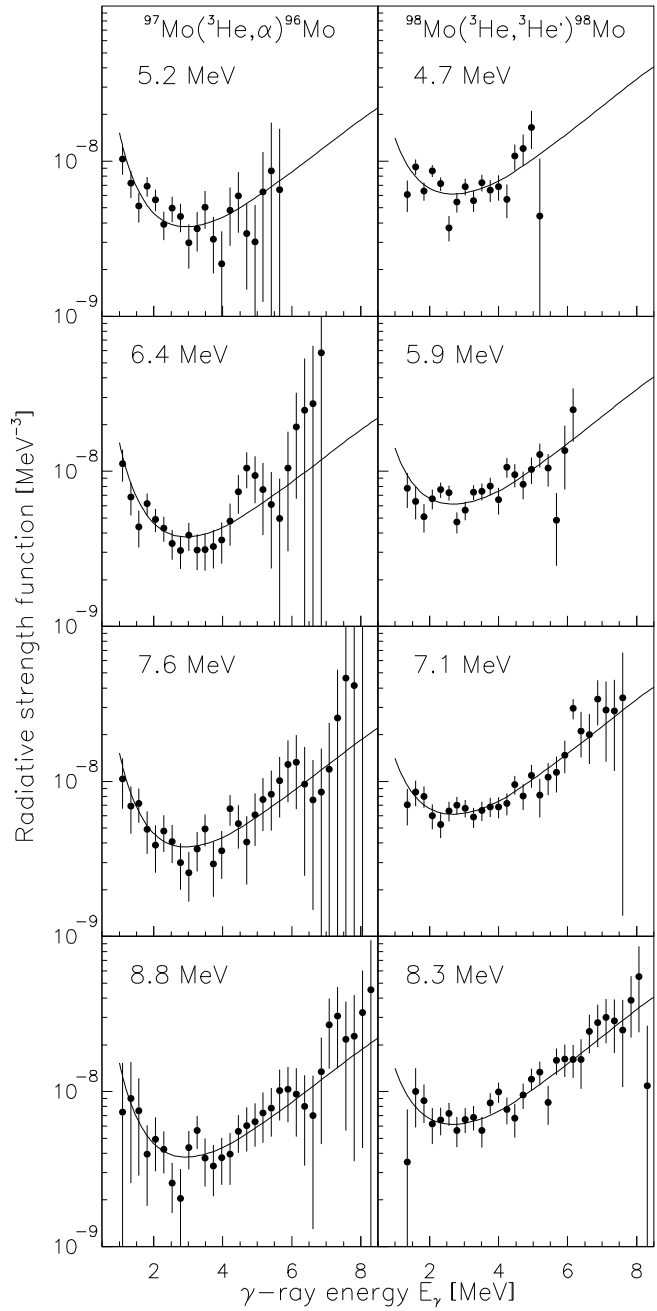}
\caption{RSFs for $^{96,98}$Mo at various initial excitation energies. The soft pole is present for all $E_i$. The solid lines display the RSFs obtained in Fig.~\ref{fig:webrsf}.}
\label{fig:rsfi68}
\end{figure}

In Fig.~\ref{fig:rsfi68}, the RSFs for $^{96,98}$Mo are shown at various initial energies $E_i$. For comparison, the figure also includes the global RSFs (solid lines) obtained with the Oslo method (Fig.~\ref{fig:webrsf}). Within the error bars the data support that the soft pole is present in all the excitation bins studied.

The origin of the soft pole cannot be explained by any known theoretical model. One would therefore
need to know the $\gamma$-ray multipolarity as guidence for theoretical approaches to this phenomenon.
Rough estimates of the reduced strength can be obtained from 
\begin{equation}
B(XL \uparrow)= \frac{1}{8\pi ^3}\frac{L[(2L+1)!!]^2}{L+1}   
\left( \frac{\hbar c}{\langle E_{\gamma} \rangle} \right)^2 I.
\end{equation}
The integrated cross-section $I$ is determined by
\begin{equation}
I= \int_{\rm 1 MeV}^{\rm 3 MeV} {\mathrm{d}}E_{\gamma} \sigma(E_{\gamma}),
\end{equation}
and the average energy is given by
\begin{equation}
\langle E_{\gamma} \rangle= I^{-1}\int_{\rm 1 MeV}^{\rm 3 MeV} {\mathrm{d}}E_{\gamma}E_{\gamma} \sigma(E_{\gamma}),
\end{equation}
where 
\begin{equation}
\sigma(E_{\gamma})=3(\pi \hbar c)^2 E_{\gamma} f_{\rm softpole}(E_{\gamma}).
\end{equation}
In the evaluation, we have integrated the soft pole between 1 and 3 MeV. Thus, the estimates 
give a lower limit for the respective $B(XL \uparrow)$ values. The correct result will of course
depend on the functional form of $f_{\rm softpole}(E_{\gamma})$ below 1 MeV, however, no
experimental data exist in this region and any assumption here would be highly speculative.

With the assumptions above, we get on the average a $B(E1)$ value of
$0.02$ e$^2$fm$^2$, which is $0.07$ \% of the sum rule for the GEDR. For $B(M1)$, we get roughly
$1.8 \mu _N ^2$, which is $3-4$ times larger than the observed strength to mixed symmetry $1^+$
states around 3 MeV \cite{frans03,frans04}. Finally, we obtain a $B(E2)$ value around $8000 $e$^2$fm$^4$, which
is $3-8$ times larger that for the excitation to the first excited $2^+$ states in the even molybdenum isotopes.
Thus, we cannot exclude any of these multipolarities, since neither of them would yield unreasonably high transition strengths.

\section{Summary and conclusions}

As expected, the observed RSFs reveal very similar shapes since they all refer to isotopes with the same nuclear charge. When going from $N=51$ to $56$ the RSF increases by almost a factor of two for $E_{\gamma}>3$ MeV, which can be understood from the change of nuclear deformation. With the onset of deformation, the increasing resonance GEDR width $\Gamma_{{\mathrm{E}}1}$ is responsible for the increasing strength.

An enhanced strength at low $\gamma$ energies is observed, which is equally strong for all isotopes and excitation energies studied. A similar enhancement has also been seen in the iron isotopes. The multipolarity of the soft pole radiation is unknown and there is still no theoretical explanation for this very interesting phenomenon.

\acknowledgements
Financial support from the Norwegian Research Council (NFR) is gratefully acknowledged. Part of this work was 
performed under the auspices of the U.S. Department of Energy by the University
of California, Lawrence Livermore National Laboratory under Contract 
W-7405-ENG-48. A.V. acknowledges support from a NATO Science Fellowship under 
project number 150027/432. 
E.A, U.A, and G.E.M acknowledge support from the National Nuclear Security
Administration under the Stewardship Science Academic Alliances program
through DOE Research Grant No. DE-FG03-03-NA00076 and U.S.
Department of Energy Grant No. DE-FG02-97-ER41042.

\end{multicols}

\newpage

\begin{table}
\caption{Parameters used for the radiative strength functions. The data are taken from Ref.~\protect\cite{IA98}. The E1 resonance parameters for the even Mo isotopes are based on photo absorption experiments \protect\cite{atom}, and the parameters for the odd Mo isotopes are derived from interpolations.}  
\begin{tabular}{l|ccc|ccc|ccc|c}
Nucleus &$E_{{\mathrm{E}}1}$&$\sigma_{{\mathrm{E}}1}$&$\Gamma_{{\mathrm{E}}1}$&$E_{{\mathrm{M}}1}$&$\sigma_{{\mathrm{M}}1}$&$\Gamma_{{\mathrm{M}}1}$&$E_{{\mathrm{E}}2}$&$\sigma_{{\mathrm{E}}2}$&$\Gamma_{{\mathrm{E}}2}$&$\langle \Gamma_\gamma \rangle$\\
          &(MeV)&(mb) &(MeV) &(MeV)&(mb)&(MeV)&(MeV)&(mb) &(MeV) &(meV) \\ \hline
&&&&&&&&&&\\
$^{93}$Mo&16.59& 173.5 &4.82&9.05& 0.86 &4.0& 13.91& 2.26 &4.99& 160(20)     \\
$^{94}$Mo&16.36& 185.0 &5.50&9.02& 1.26 &4.0& 13.86& 2.24 &4.98& 170(40)$^a$ \\
$^{95}$Mo&16.28& 185.0 &5.76&8.99& 1.38 &4.0& 13.81& 2.22 &4.97& 135(20)     \\
$^{96}$Mo&16.20& 185.0 &6.01&8.95& 1.51 &4.0& 13.76& 2.21 &4.96& 150(20)     \\
$^{97}$Mo&16.00& 187.0 &5.98&8.92& 1.58 &4.0& 13.71& 2.19 &4.95& 110(15)     \\
$^{98}$Mo&15.80& 189.0 &5.94&8.89& 1.65 &4.0& 13.66& 2.17 &4.93& 130(20)     \\
\label{tab:tab1}
\end{tabular}
\end{table}
$^a$Estimated from systematics.

\begin{table}
\caption{Parameters obtained from the fitting of the radiative strength function.} 
\begin{tabular}{l|cc|c}
Reaction&                             ${\cal{A}}$&       $b$  &  $\kappa$ \\
        &                         (mb/MeV) &          &  \\ \hline
&&&\\
($^3$He,$\alpha$)         $^{93}$Mo  &0.37(7)  &2.6(3) &0.44(4)  \\
($^3$He,$^3$He$^{\prime}$)$^{94}$Mo  &0.48(5)  &2.5(2) &0.36(2)  \\
($^3$He,$\alpha$)         $^{95}$Mo  &0.48(6)  &2.6(2) &0.39(2)  \\
($^3$He,$^3$He$^{\prime}$)$^{96}$Mo  &0.60(4)  &3.2(2) &0.36(1)  \\
($^3$He,$\alpha$)         $^{96}$Mo  &0.47(14) &2.7(6) &0.32(4)  \\
($^3$He,$^3$He$^{\prime}$)$^{97}$Mo  &0.47(7)  &2.4(3) &0.38(3)  \\
($^3$He,$\alpha$)         $^{97}$Mo  &0.30(10) &2.2(5) &0.45(5)  \\
($^3$He,$^3$He$^{\prime}$)$^{98}$Mo  &0.22(7)  &2.1(5) &0.52(4)  \\

\end{tabular}
\label{tab:tab2}
\end{table}

\clearpage

\end{document}